\documentclass[reprint,showpacs,superscriptaddres,showkeys,floatfix]{revtex4-1}
\usepackage{graphicx}
\usepackage{dcolumn}
\usepackage[latin1]{inputenc}
\usepackage{amsmath}
\usepackage{amsfonts}
\usepackage{upgreek}
\usepackage{color}
\usepackage{subfigure}

\begin{document}

\title{Spontaneous skyrmion chains in nanorods with chiral interactions}

\author{M. Charilaou, J. F. L\"offler}
\affiliation{Laboratory of Metal Physics and Technology, Department of Materials, ETH Zurich, 8093 Zurich, Switzerland}
\email{charilaou@mat.ethz.ch}

\begin{abstract}
We report that in cylindrical nanorods skyrmion chains, i.e., three-dimensional spin textures, emerge dynamically, as revealed by micromagnetic simulations. The skyrmion-chain state occurs when the diameter of the rod is larger than the helical pitch length of the material, and the number of skyrmions on the chain is proportional to the length of the nanorod. This finding provides a deeper understanding of the interplay between geometry and skyrmionic symmetry, and shows how spatial confinement can stabilize spontaneous topological spin textures.  
\end{abstract}

\maketitle

Competition between the symmetric exchange interaction, the antisymmetric Dzyaloshinskii-Moriya interaction (DMI) \cite{DMI}, and uniaxial crystalline anisotropy gives rise to the formation of complex spin textures in magnetic materials. In bulk magnets the DMI contributions typically cancel out, but non-zero net contributions emerge in low-symmetry systems, such as on surfaces and steps \cite{skomski1998,bogdanov2001,bode2007}. A fascinating manifestation of the competition between symmetric and antisymmetric interactions is the occurrence of skyrmions and skyrmion lattices \cite{skyrme1961,bogdanov1989,roessler2006}, as observed e.g. in epitaxial FeGe(111) films \cite{huang2012} and ultrathin Fe/Ir films \cite{heinze2011} with strong perpendicular anisotropy at zero field, and more recently in ultrathin Co/Pt nanostructures at room temperature \cite{moutafis2016,boulle2016}. The stability of skyrmionic states is therefore contingent upon a symmetry-breaking field, either external or internal \cite{roessler2006}, which can also be set by the geometry of a solid, i.e., the confinement of the spin structure in a low-symmetry system \cite{rohart2013}. On surfaces of bulk crystals skyrmions exist in a narrow temperature$-$field range, close to the Curie temperature \cite{muehlbauer2009,yu2010a,yu2010b}, but in thin films the skyrmion phase extends to wider temperature- and field-ranges \cite{yu2010a,huang2012}. The confinement of spin textures in nanostructures is therefore a key element for the stability of these states.

The intense research on skyrmions is fueled on the one hand by the fascination of the new physics related to these complex spin structures, and on the other hand by the potential to develop new technology for data-storage devices. This is motivated by the fact that skyrmions can be moved by relatively low current densities
\cite{jonietz2010,iwasaki2013,fert2013a,fert2013b,zhou2014a}, promising energy-efficient spintronics, and because skyrmion-based devices can be coupled to conventional domain-wall-based devices by adjusting the geometry of the solid \cite{zhou2014b}.  

The confinement of skyrmionic spin textures in nano-structures is a key element in creating and controlling them, and as we will discuss in this letter, the geometry of nanostructures can give rise to complex spontaneous spin textures. Using high-resolution micromagnetic finite-difference simulations for the structure of FeGe we will show that in cylindrical nanorods three-dimensional metastable topological spin textures emerge, which exhibit a one-dimensional oscillation of the topological charge. First we will discuss simulations of ultrathin nanodisks, and then we will show results for cylindrical nanorods.

In our calculations, the total energy density $F$ is the sum of the ferromagnetic exchange 
\begin{equation}
F_\mathrm{exc} = A_\mathrm{exc} \left(\nabla \mathbf{m}\right)^2,
\end{equation}

\noindent where $A_\mathrm{exc}$ is the exchange stiffness and $\mathbf{m}$ is the magnetization unit vector ($\mathbf{m}=\mathbf{M}/M_\mathrm{S}$ with $M_\mathrm{S}$ the saturation magnetization); the Dzyaloshinskii-Moriya interaction 

\begin{equation}
F_\mathrm{DMI} = D \mathbf{m} \cdot \left( \nabla \times \mathbf{m} \right), 
\end{equation}

\noindent where $D$ is the strength of the DMI; the Zeeman energy 
\begin{equation}
F_\mathrm{Z}=\mu_0 \mathbf{H}_\mathrm{ext}\cdot \mathbf{m},
\end{equation}

\noindent where $\mathbf{H}_\mathrm{ext}$ is the external field; and the magnetostatic self-energy due to dipolar-interactions 
\begin{equation}
F_\mathrm{dip}=-\frac{\mu_0}{2}\mathbf{m} \cdot \mathbf{h}_\mathrm{demag},
\end{equation}

\noindent where $\mathbf{h}_\mathrm{demag}$ is the local demagnetizing field.  

For the simulations we used the graphics-processing-unit accelerated software package \textit{MuMax3} \cite{mumax3}, which solves the Landau-Lifshitz-Gilbert equation of motion 

\begin{equation}
\partial_t  \mathbf{m}= -\gamma \left(\mathbf{m} \times \mathbf{h}_\mathrm{eff}\right)+\alpha \left(\mathbf{m} \times \partial_t \mathbf{m}\right),
\end{equation} 

\noindent where $\alpha=0.1$ is the dimensionless damping parameter ($\alpha=G/\gamma M_\mathrm{S}$, with $G$ the Gilbert damping frequency constant and $\gamma$ the electron gyromagnetic ratio), and $\mathbf{h}_\mathrm{eff}=-(1/\mu_0 M_\mathrm{S})\partial_\mathbf{m} F$ is the effective magnetic field consisting of both internal and external fields. The material parameters for FeGe taken from literature are: exchange stiffness \cite{beg2014} $A_\mathrm{exc}=8.78$~pJ/m, saturation magnetization \cite{ericsson1981,yamada2003} $M_\mathrm{S}=385$~kA/m, and DMI strength \cite{beg2014} $D=1.58$~mJ/m$^2$. Since the DMI energy in FeGe is intrinsic and not of interfacial origin, it should not depend on the thickness, hence we kept the DMI strength constant throughout all simulations. 

The observables recorded during the simulations were i) the local magnetic moment configuration in the solid $\mathbf{m}$, ii) the net magnetization $\mathbf{M}$, and iii) the 2D topological charge $Q$  
\begin{equation}
Q=\frac{1}{4\pi} \int \mathbf{m} \cdot \left( \partial_x \mathbf{m} \times \partial_y \mathbf{m} \right) \mathrm{d}x \mathrm{d}y.
\end{equation}

The cell size for the simulations of the nanodisks was set to 0.5 nm x 0.5 nm x 0.5 nm, and for the nanorods it was varied to achieve maximum performance while maintaining high spatial resolution. For the image shown in Fig. \ref{fig2} the cell size was 3.75 nm x 3.75 nm x 3.9 nm, whereas for more detailed analysis we used 1.875 nm x 1.875 nm x 1.97 nm (mesh of $64\times 64\times 256$ for a rod with diameter of 120 nm and length of 500 nm). Occasional checks were made with smaller cell sizes to confirm the validity of the results.

\begin{figure}
	\centering
		\includegraphics[width=1.0\columnwidth]{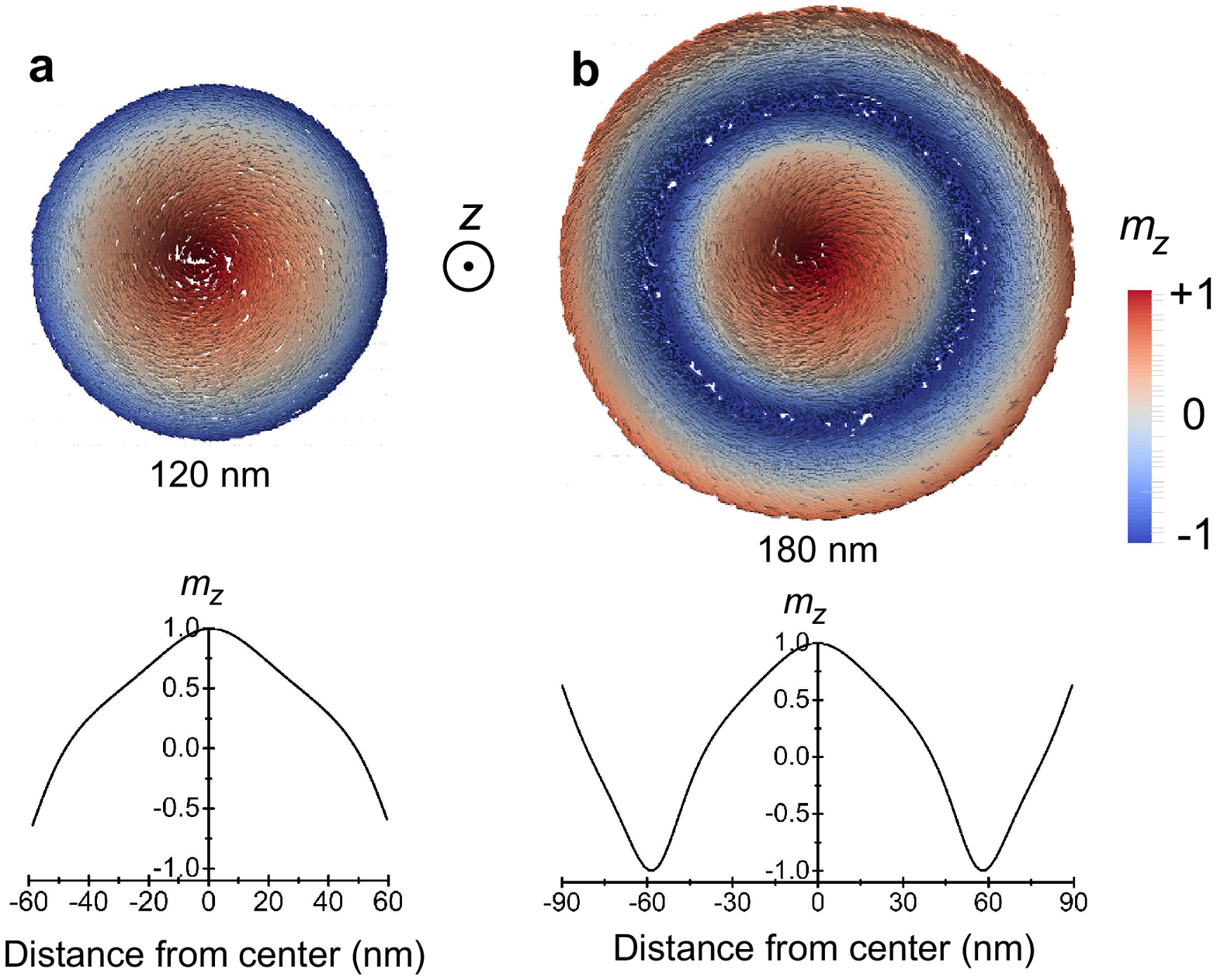}
	\caption{\textbf{2D confined skyrmions.} Ultrathin nanodisks with diameter of (a) 120 nm and (b) 180 nm, where a skyrmion core is geometrically confined. The plots below each disk show the magnetization profile, i.e., the $z$-component of the magnetization unit vector as a function of the distance from the disk center. }
	\label{fig1}
\end{figure}

 \begin{figure*}
	\centering
		\includegraphics[width=2.0\columnwidth]{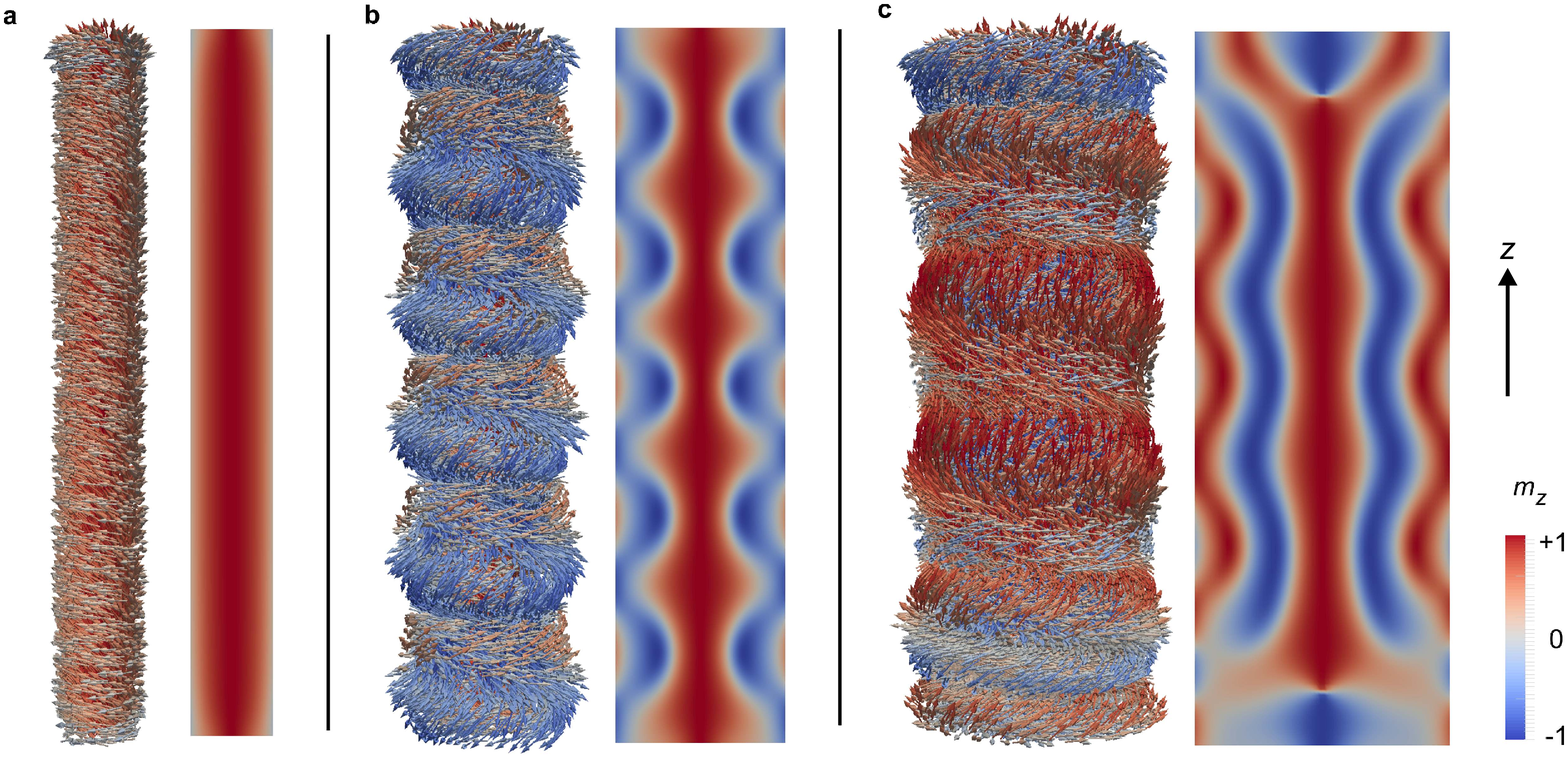}
	\caption{\textbf{Skyrmion chains in nanorods.} Magnetic configurations in cylindrical nanostructures (nanorods) with a length of 500 nm and a diameter of (a) 60 nm, (b) 120 nm, and (c) 180 nm. For each rod the spin structure is shown as a vector field (left) and a contour plot of the magnetization along the $z$ axis (right).  }
	\label{fig2}
\end{figure*}

\begin{figure}
	\centering
		\includegraphics[width=1.0\columnwidth]{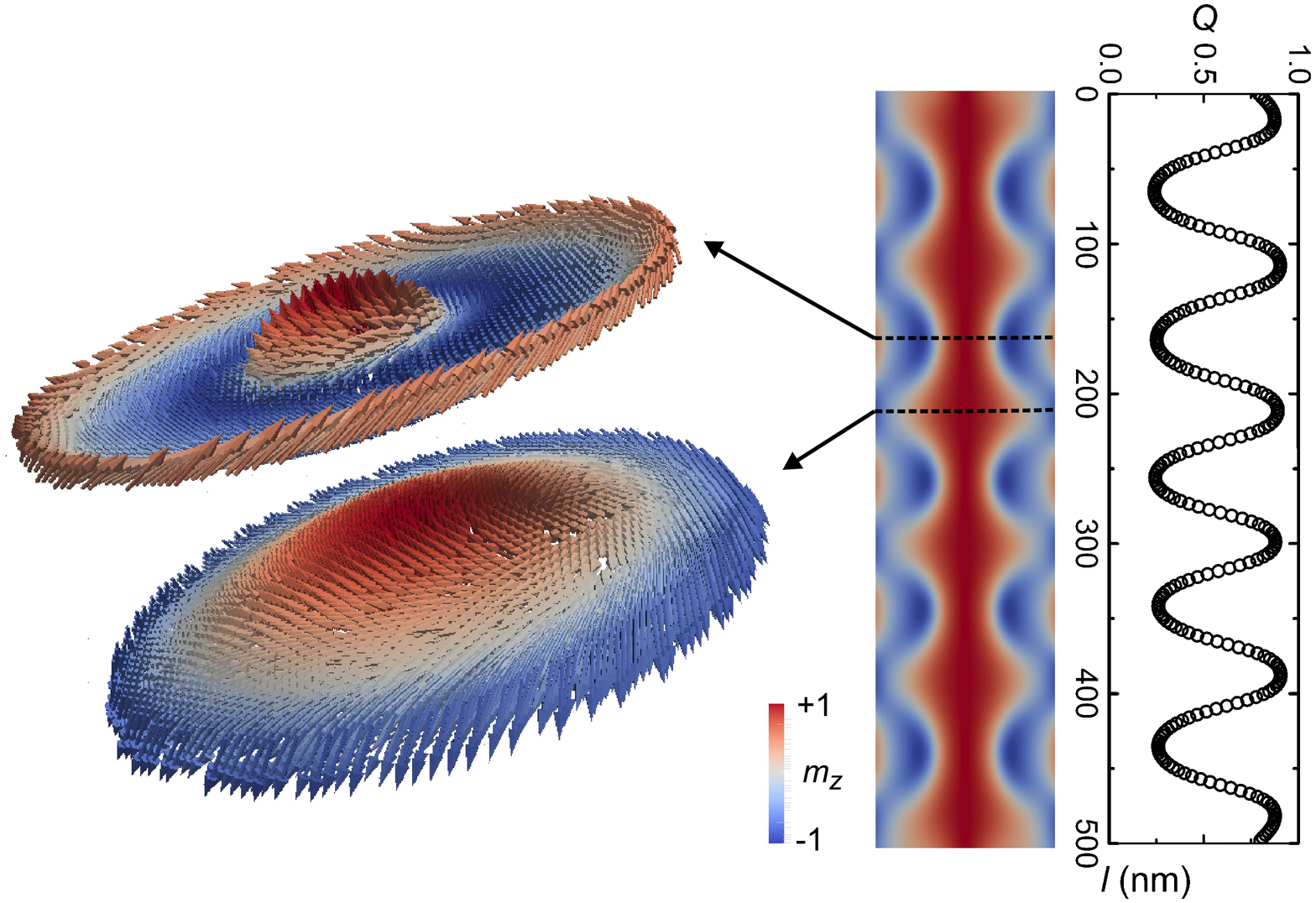}
	\caption{\textbf{Oscillation of the topological charge.} Cross-section of a nanorod with 120 nm diameter and 500 nm length showing the oscillation of the spin structure and the topological charge $Q$ as a function of the distance ($l$) from the rod end. The two cross-sections show the skyrmion (high $Q$) and mexican hat (low $Q$) spin configuration. }
	\label{fig3}
\end{figure}

A crucial aspect for the realization of a skyrmion in an ultrathin structure, e.g. a nanodisk, is the presence of a symmetry-breaking field, either external or internal. It also possible to prepare the skyrmion state by applying external fields, as shown for Co/Pd nanodots \cite{liu2015}. In our simulations we can dynamically form a metastable skyrmion even in the absence of uniaxial anisotropy by magnetizing the sample in the out-of-plane direction using a field pulse ($\mu_0 H_z=1$ T) and allowing the spin structure to relax. Doing this results in a left-handed skyrmion core (see Fig. \ref{fig1}a), in which the magnetization in the center of the disk is $m_z=+1$ and around the perimeter of the disk is $m_z\approx -1$. We quantify the skyrmionic configuration by calculating the topological charge $Q$, and find a numerical value of 0.8 (for a perfect skyrmion of $Q=\pm 1$). The deviation from the integer value is due to the tilting of the spins along the circumference of the disk by dipolar interactions.

For the confinement of the skyrmionic spin texture, the size of the disk needs to be comparable to the skyrmion-core radius, which depends on the interplay between the DMI and magnetostatic interactions \cite{rohart2013}. Figure \ref{fig1} shows two simulation results for diameters of (a) 120 nm and (b) 180 nm along with the magnetization profiles as a function of the distance from the disk center. For 120 nm the result is a complete skyrmion core, and for 180 nm the spin structure forms a concentric ring with alternating $m_z$, similar to those shown in Refs. \cite{moutafis2007,rohart2013,beg2014}. If we further increase the radius of the disk, the ring configuration breaks down to a helical state (not shown here). We find that a single skyrmion in FeGe ultrathin disks is stable for diameters in the range of 70 nm -- 135 nm, i.e., $\lambda < d < 2 \lambda$, where $\lambda = 2\pi A_\mathrm{exc}/D\approx 70$~nm is the characteristic pitch length for FeGe ($A_\mathrm{exc}$ is the ferromagnetic exchange energy and $D$ is the strength of the DMI). Note that without considering the energy contribution of dipolar interactions ($F_\mathrm{dip}$), a single skyrmion is only stable up to $d\leq 90$ nm. Dipolar interactions are therefore crucial in stabilizing the skyrmion, as they tend to align the spins along the physical edge of the disk, thus shrinking or stretching the skyrmion in order to satisfy this condition. This importance of dipolar interactions was recently also discussed with regards to experiments on Pt/Co/MgO nanostructures \cite{boulle2016}, where it was found that with increasing DMI strength the role played by dipolar interactions became increasingly important. This has to do with the interplay between two length scales, i.e., $\lambda$ and the exchange length \cite{braun2012} $\delta_M=2\sqrt{A_\mathrm{exc}/(\mu_0 M_\mathrm{S}^2)}\approx 14$~nm for FeGe. When $\lambda>>\delta_M$, the curling period of the spin structure is longer than the exchange length. As, however, the DMI increases and $\lambda$ decreases, the curling is impeded by the dipolar interactions and the role of magnetostatics becomes more important.  

\begin{figure*}
	\centering
		\includegraphics[width=2.0\columnwidth]{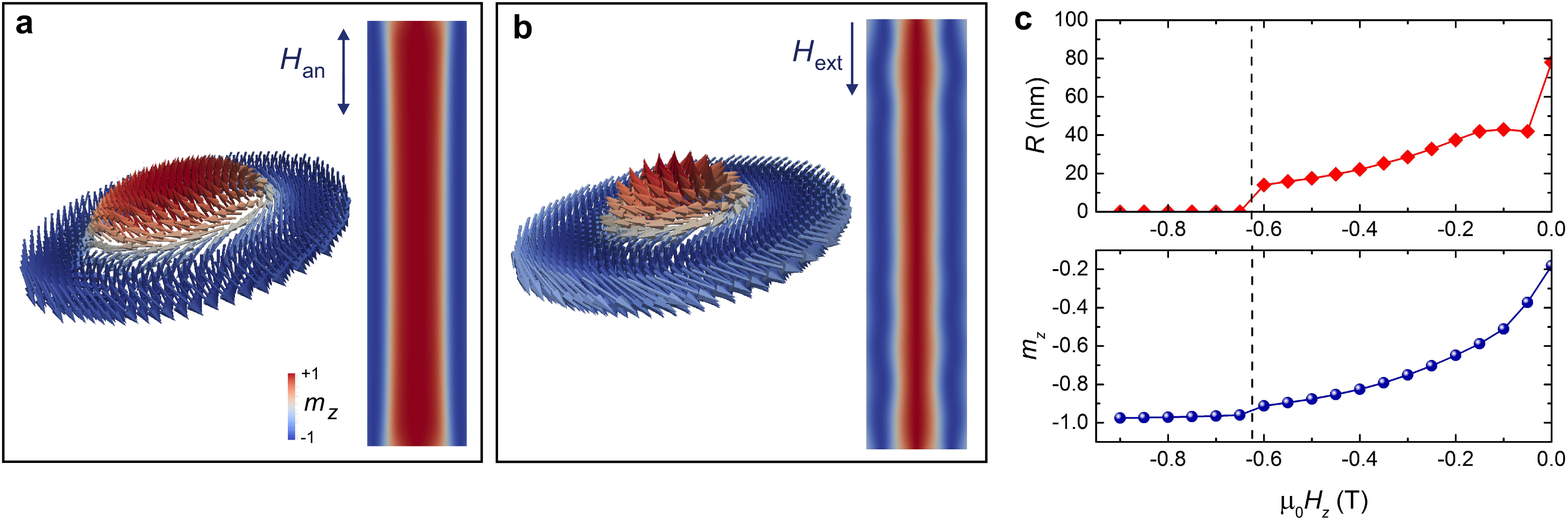}
	\caption{\textbf{Symmetry breaking and skyrmion lines.} Contour plots of the magnetization in a nanorod with diameter 120 nm and length 500 nm, (a) where there is a uniaxial anisotropy field, and (b) when we apply an external field. The spin structure corresponds to that of a single skyrmion line. (c) With increasing opposing external field, the skyrmion radius ($R$) decreases and the skyrmion line vanishes at the critical field where $m_z=-1$. }
	\label{fig4}
\end{figure*}

Now we turn to 3D structures in order to examine the dimensional evolution of the skyrmion state in cylindrical symmetry. The state was prepared in a similar was as for the nanodisks, i.e., saturating external field applied along the $z$ axis and then removed. In a nanorod with a diameter of 60 nm the spin configuration corresponds to an incomplete helix (see Fig. \ref{fig2}a), where the core of the rod is magnetized along the $z$ axis and the spins on the circumference of the rod are left-handedly wrapped around the solid. If we increase the diameter to 120 nm, however, corresponding to a diameter at which the skyrmion core is complete in an ultrathin nanodisk, the spin structure is a striking three-dimensional \textit{skyrmion chain}, where the core of the rod is magnetized along the $z$ axis but the spins away from the core form a three-dimensional spin texture with a spatial oscillation along the rod (see Fig. \ref{fig2}b).

We investigated a range of diameters and found that the stability range for a skyrmion chain in a rod is 90 nm $<d<$ 135 nm. For larger diameters, i.e., for $d=180$ nm, the spin structure is in a complex helical-like state (see Fig. \ref{fig2}c), where the variation of the moments from one edge to the other is discontinuous, and unfolds via the formation of hedgehogs, or Bloch points, close to the edges of the rod, similar to those shown by Milde \textit{et al.} \cite{milde2013}.  

Increasing the rod length increases the number of skyrmions in the chain. The dipolar interactions play an important role here, as they stretch or compress the skyrmions in order to adapt the spin texture onto the solid and fulfill the boundary conditions, as discussed above. Importantly, the number of skyrmions in the rod can only be increased by integer steps: when having a rod of certain length that contains 3 skyrmions, the number of skyrmions will remain the same while we increase the length up to the point where the rod can fit 4 skyrmions, etc. This scenario of skyrmion packing is topologically equivalent to atomic cluster-packing, where the number of nearest neighbors around a solute atom remains constant upon increasing the radius of the atom, and then jumps to a new configuration above a threshold value \cite{miracle2004}. 

For a nanorod with a diameter of 120 nm and length of 500 nm, the spin structure contains 6 skyrmions, as seen in Fig. \ref{fig3}a, which shows the oscillation of $Q$ inside the solid. The oscillation of the spin structure from one skyrmion core, i.e., where $Q\approx 0.9$, to the next is done via a helical mexican-hat-like spin texture (see Fig. \ref{fig3}). The oscillation of $Q$ has a sinusoidal form $Q\propto \sin(\pi l/\Lambda)$, with a period of $\Lambda=95$~nm. Importantly, while the spin texture seems continuous, and the skyrmions along the chain connected, we see that the topological charge is localized at the skyrmion configuration (see Fig. \ref{fig3}).

The localization of $Q$ can be suppressed by a symmetry-breaking field, either internal or external. Let us consider a hypothetical scenario of a material with exactly the same parameters ($M_\mathrm{S}$, $A_\mathrm{exc}$, and $D$), which additionally has a uniaxial magnetocrystalline anisotropy $K_u$ that is comparable to the magnetostatic self-energy, i.e., $K_u=\mu_0 M_\mathrm{S}^2/2\approx 10^5$~J/m$^3$. As shown in Fig. \ref{fig4}a, the 3D oscillation of the spin structure vanishes and the result is a continuous \textit{skyrmion line} along the rod with $Q=0.85$. Similarly, if we break the symmetry by an external field along the $z$ direction opposing the magnetization in the core (without having $K_u$), the resulting spin configuration is again a single skyrmion line along the rod. For a one-to-one comparison between the effect of internal vs. external field, the applied field in this example was set equal to the anisotropy field from the example shown in Fig. \ref{fig4}a, i.e., ($\left|\mu_0 H_z\right|=\mu_0 H_\mathrm{an}=2K_u/M_\mathrm{S}=0.52$~T). The external field not only generates a single skyrmion line, but also decreases the skyrmion radius (see Fig. \ref{fig4}c). In fact, with increasing (opposing) $H_z$, the skyrmion radius decreases monotonically up to the critical field, at which $Q\rightarrow 0$ and $m_z\rightarrow -1$. Once $m_z=-1$, if we switch off the external field, the spin configuration will return to that of a skyrmion chain, but with opposite polarity.

All the predictions made here may be verified experimentally, either by real-space observation, i.e., Lorentz transmission electron microscopy or magnetic force microscopy, or by reciprocal space investigations, such as polarized small-angle neutron scattering.

In summary, we have shown that geometrical confinement enables the occurrence of spontaneous three-dimensional skyrmion chains in cylindrical nanorods, and that a symmetry-breaking field, either external or internal, turns the structure to that of a single skyrmion line. These findings provide a deeper understanding of the stability of skyrmionic spin configurations in nano-structures, where spatial confinement plays a vital role. They may also be of great importance for the further development of spintronics towards skyrmion-based technologies.

\section*{Acknowledgment}
The authors thank H.-B. Braun for fruitful discussions. We gratefully acknowledge funding from the Swiss National Science Foundation.

\end{document}